\documentclass[aps,pra,reprint,superscriptaddress,floatfix]{revtex4-2}
\usepackage[T1]{fontenc}
\usepackage[utf8]{inputenc}
\usepackage{lmodern}
\usepackage{amsmath,amssymb,bm}
\usepackage{graphicx,xcolor}
\usepackage{subfigure}
\usepackage[colorlinks=true,linkcolor=blue!50!black,citecolor=blue!50!black,urlcolor=blue!50!black]{hyperref}

\begin{document}
\title{Ultrahigh-speed digital holography reveals blood flow beneath the skin}
\author{Yann Fischer}
\affiliation{Institut Langevin, CNRS, ESPCI Paris, PSL University, Paris, France}
\author{Zacharie Auray}
\affiliation{Institut Langevin, CNRS, ESPCI Paris, PSL University, Paris, France}
\author{Maxime Boy-Arnould}
\affiliation{Institut Langevin, CNRS, ESPCI Paris, PSL University, Paris, France}
\author{Michael Atlan}
\affiliation{Institut Langevin, CNRS, ESPCI Paris, PSL University, Paris, France}
\date{11 November 2026}
\begin{abstract}
Laser Doppler holography (LDH) leverages ultrahigh-speed digital holography to reveal blood-flow signals in vessels beneath the skin of the closed eyelid and around the eye. We demonstrate this in one volunteer using recordings acquired at approximately 16,000 frames/s under 10~mW of diffuse near-infrared illumination. Numerical Fresnel reconstruction and principal-component filtering reveal branching vascular patterns. These results extend an ocular imaging system to blood-flow visualization beneath the skin without requiring elaborate immobilization procedures.
\end{abstract}
\maketitle

\begin{figure*}[!t]
 \centering
 \subfigure[]{\begin{minipage}[b]{0.46\textwidth}
  \includegraphics[width=0.49\linewidth]{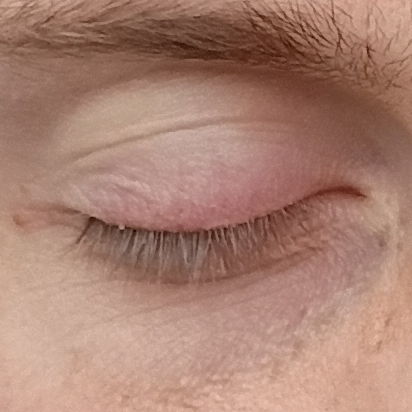}\hfill%
  \includegraphics[width=0.49\linewidth]{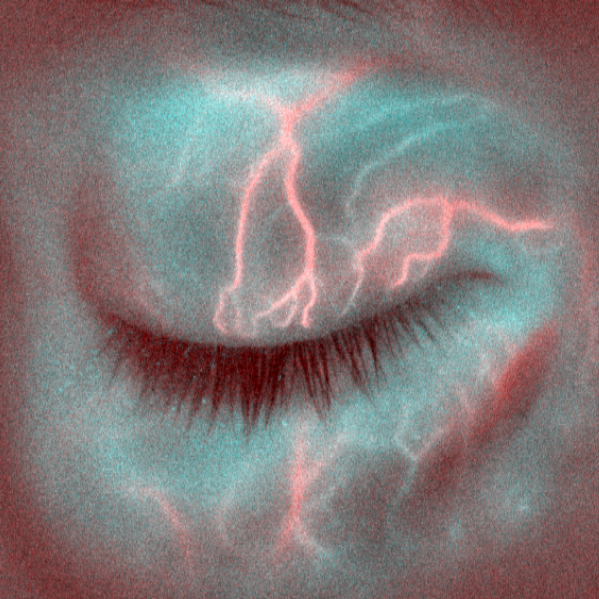}
 \end{minipage}}\hfill%
 \subfigure[]{\begin{minipage}[b]{0.46\textwidth}
  \includegraphics[width=0.49\linewidth]{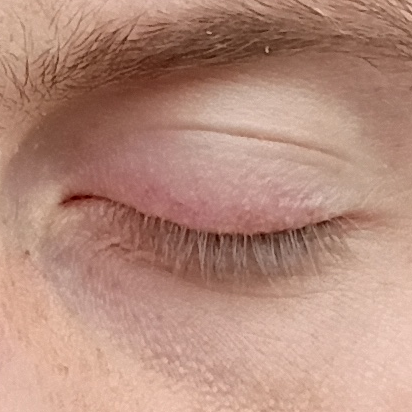}\hfill%
  \includegraphics[width=0.49\linewidth]{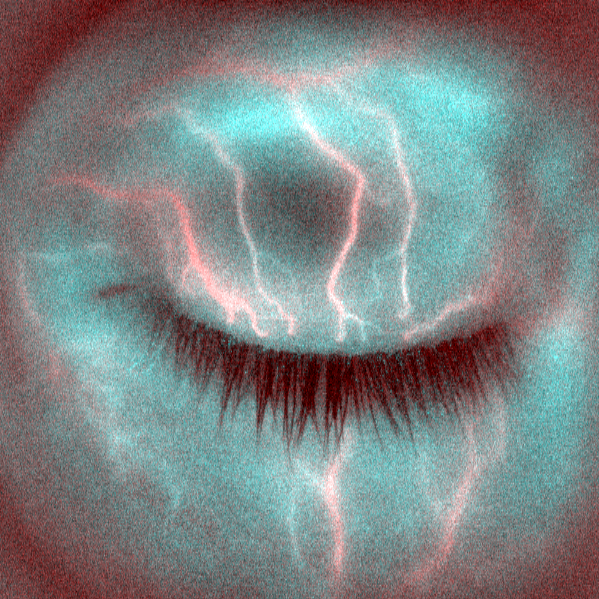}
 \end{minipage}}\par\smallskip
 \subfigure[]{\begin{minipage}[b]{0.46\textwidth}
  \includegraphics[width=0.49\linewidth]{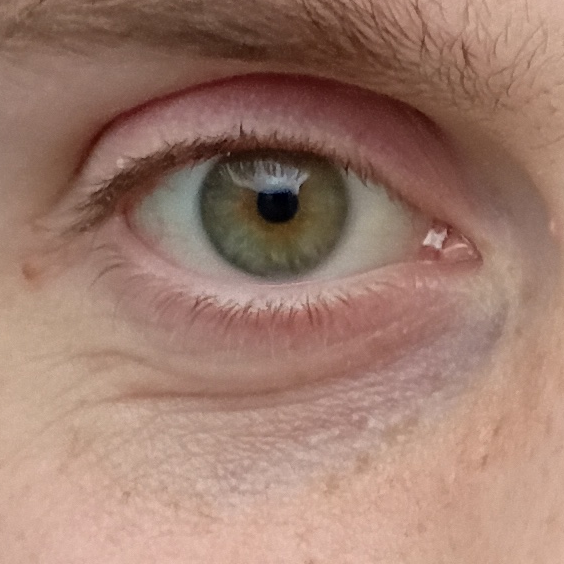}\hfill%
  \includegraphics[width=0.49\linewidth]{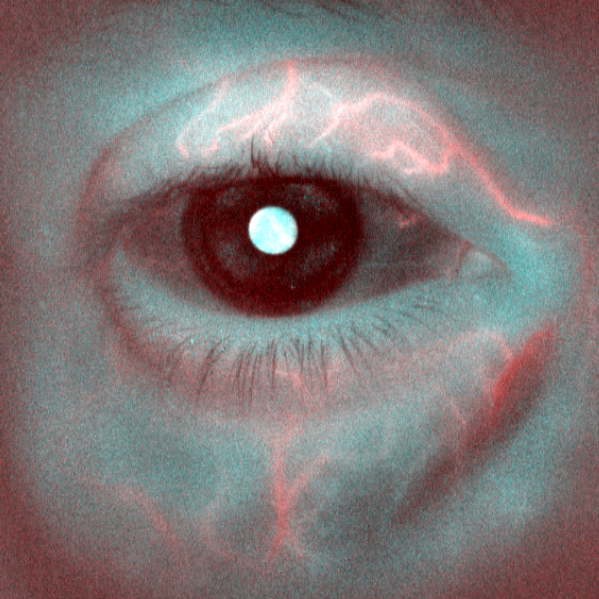}
 \end{minipage}}\hfill%
 \subfigure[]{\begin{minipage}[b]{0.46\textwidth}
  \includegraphics[width=0.49\linewidth]{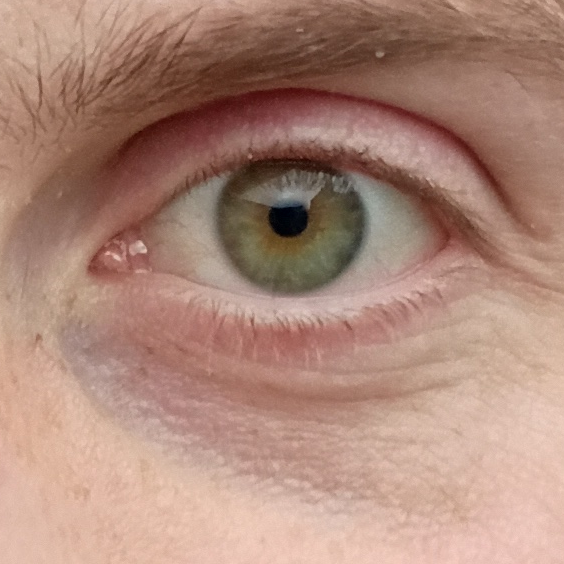}\hfill%
  \includegraphics[width=0.49\linewidth]{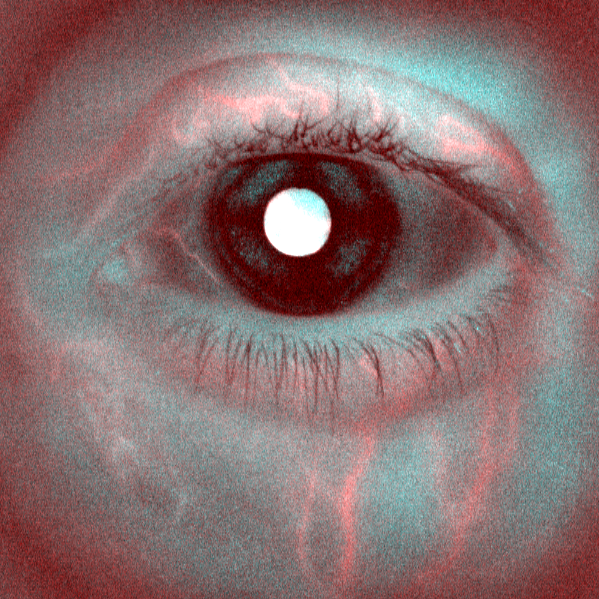}
 \end{minipage}}
 \caption{Paired photographs and LDH views of one volunteer: (a,b) right and left eyes closed; (c,d) right and left eyes open. Photographs orient the eyelid and periocular anatomy; LDH reveals branching vascular contrast over the closed lids and around the open eyes, together with a bright pupil return. Crops provide approximate anatomical correspondence without image registration. Red and cyan retain the 8 and 28 smallest-singular-value components, respectively.}
 \label{fig:eyes}
\end{figure*}

\par\medskip
\noindent\textbf{Introduction.}
Blood-flow dynamics beneath the skin complement visible vascular appearance. Laser Doppler methods detect fluctuations associated with moving blood cells~\cite{Bonner1981}; parallel instruments and laser speckle contrast imaging established wide-field cutaneous measurements~\cite{Serov2002,Serov2005,FullField2011,Briers1996,Briers1996Reconciliation}. Multi-exposure and high-speed methods address static scattering and resolve changing perfusion~\cite{Parthasarathy2008,Rivera2023,Hultman2018MELSCI,Hultman2020MELSCI}. Spatially depolarized dynamic speckle imaging and full-field amplitude speckle decorrelation angiography further delineate cutaneous microvascular networks using polarization filtering and complementary speckle contrast and decorrelation metrics~\cite{Colin2022DepolarizedSpeckle,Mansutti2024FASDA}. Human studies demonstrate responses to vascular provocations and applications in systemic sclerosis~\cite{Puissant2013LSCI,Roustit2010LSCI,Ruaro2013LASCA}, while emphasizing the dependence of perfusion indices on tissue optics and spatial and temporal averaging~\cite{Millet2011LSCI,Rousseau2011ROI,SkinMESI2023Calibration}.

Holographic imaging through scattering media began with conventional wavefront reconstruction through inhomogeneous and diffusing media in the 1960s~\cite{Kogelnik1965Inhomogeneous,Leith1966Diffusing,Goodman1966Random,Stetson1967Fog,Kogelnik1968Random}, followed by reduced-coherence approaches for scattering media and biological tissue~\cite{Leith1992Scattering}. Digital methods introduced wavelength synthesis and coherence gating~\cite{Arons1995Fourier,Indebetouw2000Gating}, as well as optical phase conjugation through strongly scattering media~\cite{Hillman2013DOPC}. Li and Zhong subsequently demonstrated dynamic imaging through turbid media using continuous-wave digital holography, numerical autofocus, and spatial filtering~\cite{Li2014Turbid}. This history motivates computational access to obscured structures; the present approach uses temporal fluctuations of scattered light to reveal vascular contrast beneath skin.

Digital holography records the complex optical field through interference with a reference beam and enables numerical refocusing~\cite{Schnars1994,Gross2007}. Laser Doppler holography (LDH) combines ultrahigh-speed digital holography with Doppler contrast to visualize blood-flow dynamics. Heterodyne and short-time spectral methods established parallel Doppler imaging~\cite{Atlan2006Frequency,Atlan2006Revisited,Samson2013,Magnain2014}, followed by retinal LDH and spatiotemporal filtering~\cite{Puyo2018,Puyo2020SVD}. Earlier heterodyne holography also imaged pulse-induced out-of-plane skin motion on the thumb during an occlusion--reperfusion experiment~\cite{Bencteux2015}. Similarly, LDH has previously revealed choroidal vessels beneath the retinal pigment epithelium in the eye fundus, enabling investigations of choroidal vascular anatomy, blood-flow waveforms, and vascularity~\cite{Puyo2019Choroid,Puyo2019Waveform,Puyo2020ReverseContrast,Paques2025Submacular,Barre2026ChoroidalArteries,Auray2025ChoroidalVascularity}. Here, paired photographs and LDH images of both eyes of one volunteer show branching vascular contrast across the closed lids and surrounding periocular skin. The contribution is a direct visual demonstration of the skin field accessible with an ocular LDH platform, supported by openly available recordings and media visualizations.

\par
\begin{figure}[t]
 \centering
 \includegraphics[width=\linewidth]{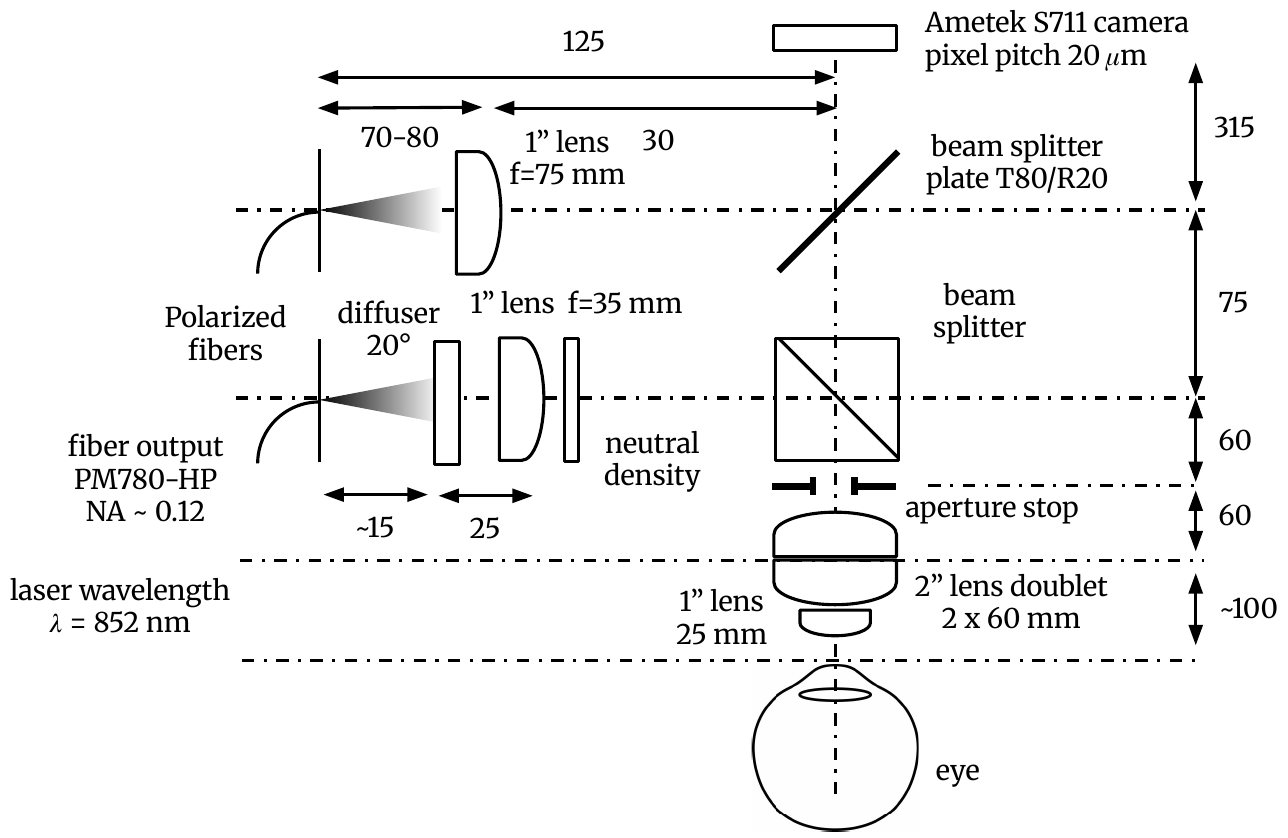}
 \caption{Optical layout of the 852-nm on-axis LDH instrument adapted for eyelid and periocular imaging. Fiber-delivered illumination passes through a diffuser and relay optics toward the eye; backscattered light is combined with the reference field and recorded by the camera. The diagram identifies the beam splitters, neutral-density filter, aperture stop, and relay lenses. Indicated distances and focal lengths are in millimeters. The eye schematic marks the measurement region.}
 \label{fig:setup}
\end{figure}

\medskip
\noindent\textbf{Methods.}
The on-axis LDH prototype used for eyelid and periocular imaging is shown in Fig.~\ref{fig:setup}. The optical configuration derives from an inline Mach--Zehnder digital holography ophthalmoscope operating at $\lambda_0 \sim 852$~nm~\cite{Auray2025UltrahighSpeed}. Light from a diode laser (Thorlabs FPV852P) was divided between the illumination and reference arms in a 90:10 power ratio using polarization-maintaining fibers. The illumination arm contained an engineered diffuser (Thorlabs ED1-C20-MD, $20^\circ$ circular top-hat) and three biconvex relay lenses with focal lengths of 60, 60, and 25~mm. This illumination arm builds on diffuse Maxwellian retinal illumination~\cite{Bratasz2022Diffuse}, also used in high-speed retinal speckle imaging~\cite{Auray2026RetinalLSCI}; here, the diffuse field illuminated the eyelid and surrounding facial skin. Cross-polarized backscattered light from the imaged region was recombined with the reference field and recorded interferometrically with a Phantom S711 camera (Ametek Vision Research) over $768\times512$ pixels with a $d=20$~$\mu$m pixel pitch.

The detection numerical aperture (NA) subtended at the eyepiece is small. In air, the paraxial sensor limit is $\mathrm{NA}_{\mathrm{sensor}}\simeq Nd/(2L)$: with a sensor--eyepiece distance $L\sim0.50$~m, this gives approximately 0.015 ($N=768$) horizontally and 0.010 ($N=512$) vertically. The on-axis fringe-sampling limit, $\mathrm{NA}_{\mathrm{sampling}}=\lambda_0/(2d)\simeq0.021$ at 852~nm, is less restrictive. The reconstructed pixel pitch is $p\sim100$~$\mu$m. 
The object-side numerical aperture at the skin was estimated by back-propagating the sensor acceptance cone through the relay optics. Approximating the three adjacent lenses ($f=25$, 60, and 60~mm) as thin lenses gives an effective focal length $f_{\mathrm{eff}}\simeq13.6$~mm. For a skin-relay distance $\ell \simeq10$~cm, we have
$\mathrm{NA}_{\mathrm{skin}}\simeq Nd/(2|\ell+L-\ell L/f_{\mathrm{eff}}|)$,
where $Nd/2$ is the sensor half-width along the considered axis. The resulting sensor-limited object-side NA at the field center is approximately 0.0025 horizontally and 0.0017 vertically. The $\sim$ 25~mm relay aperture does not restrict this acceptance. The estimated object-side NA is substantially smaller than the sensor-side NA (0.015/0.010) because of the relay geometry. The corresponding diffraction-limited lateral resolution scale, using \(0.61\lambda_0/\mathrm{NA}\), is approximately 208 $\mu$m horizontally and 306 $\mu$m vertically.

\par\medskip
\noindent\textbf{Recording conditions.}
Both eyes of a healthy 25-year-old man were imaged with the lids closed and open after written informed consent. The procedures adhered to the Declaration of Helsinki. The optical power at the diffused focal spot was 10~mW, distributed over approximately $5\times5$~cm$^2$ at the imaged region. Each recording comprised 65,536 interferograms acquired at approximately 16,000 frames/s (approximately 62~$\mu$s frame period), with a 58~$\mu$s exposure, for a duration of approximately 4~s. Fresnel propagation used a numerical reconstruction distance of 508~mm. Processing used Holovibes v14.14.1~\cite{Dubosc2025Holovibes,Holovibes14141}.

Principal-component analysis (PCA)~\cite{Puyo2020PCA} used 32-frame windows with a stride of 16 frames. Components associated with the 8 smallest singular values were retained for the red channel, and those associated with the 28 smallest singular values for the cyan channel. With singular values ordered from largest to smallest, these are components 25--32 and 5--32, respectively, using one-based indexing. Sixteen successive PCA outputs were averaged in a rolling buffer before color-channel intensity mapping. At the nominal acquisition rate, each PCA window spans 2~ms and advances by 1~ms; the 16 consecutive windows contributing to an averaged image have a combined input support of approximately 17~ms. Normalization and display gain were adaptive for all four LDH views and associated videos; displayed intensities do not provide a fixed quantitative scale across views or time. Red and cyan encode different retained subspaces; their relative intensities reflect Doppler broadening.

Conventional photographs (Fig.~\ref{fig:eyes}) were cropped to match the eye and surrounding skin in each LDH view. They provide anatomical orientation and identify the eyebrows, lid margins, eyelashes, canthi, and infraorbital skin.

\par\medskip
\noindent\textbf{Results.}
The closed-eye views reveal distinct branching vascular patterns across both upper lids (Fig.~\ref{fig:eyes}a,b). Over the right lid, a prominent central branch extends upward from near the lash line and divides toward the upper field, with curved branches extending toward the inner corner. Over the left lid, several elongated, roughly vertical branches descend toward the lid margin and connect with curved segments across the upper lid. The paired photographs locate these networks relative to the lid fold and eyelashes, making the relationship between surface appearance and LDH vascular contrast directly visible.

With the eyes open (Fig.~\ref{fig:eyes}c,d), vascular contrast remains visible in the skin above and around the eyes. A curved vessel is prominent above the right upper lid, with further contrast toward the inner canthus and lower periocular skin; the left-eye view shows finer vascular features above the lid and in the surrounding skin. Both open-eye views contain a bright return within the pupil, spatially distinct from the periocular vascular field. Doppler-shifted fundus light can contribute to such a pupil signal~\cite{Puyo2021Anterior}. Together, the four pairs demonstrate vascular visibility across eyelid states. Eye opening and closure are viewing conditions in this comparison, rather than a controlled vascular intervention; no physiological response is inferred from their differences. Associated media visualizations are available with the raw dataset~\cite{SkinRawDataset2026}.

\par\medskip
\noindent\textbf{Clinical outlook.}
Periocular vascular monitoring could complement surface inspection during reconstruction and inflammatory disease follow-up. Existing optical studies have monitored eyelid graft revascularization, flap perfusion, and responses to local anesthetics~\cite{Berggren2021Graft,Tenland2021Bilamellar,Berggren2020Flap,Tenland2025Stretch,Wiktorin2022,Sheikh2018,Bunke2022}. Lid-margin angiography and studies of intense pulsed light provide related precedents in meibomian gland dysfunction and ocular rosacea~\cite{LidOCTA2024,IPL2025RCT,Hoarau2025,IPL2025Predictors}. Longitudinal wound and vascular-lesion imaging further motivates following vascular patterns alongside tissue appearance~\cite{Vennstrom2025Healing,Stridh2024}; reported choroidal responses to periocular warming also motivate investigating superficial and intraocular signals together~\cite{Ujita2024Warming}. LDH could test whether changes within individual visible vessels add useful information to these assessments.

\par\medskip
\noindent\textbf{Limitations.}
This single-participant demonstration provides qualitative flow-sensitive images, without absolute flow quantification, depth localization, or evidence of treatment response. Numerical refocusing does not establish vessel depth; OCT or photoacoustic imaging could provide complementary localization~\cite{Wang2010SkinOCT,Zhang2006Photoacoustic}. Skin displacement and eye movement can contribute to coherent-light fluctuations~\cite{Bencteux2015,Chizari2020}, and PCA may suppress both motion and vascular signals. Quantitative interpretation therefore requires motion controls, analysis of signals before adaptive display mapping, and repeated measurements after repositioning~\cite{BlandAltman1986}.

\par\medskip
\noindent\textbf{Conclusion.}
LDH reveals branching vascular contrast beneath the closed eyelid and in the skin surrounding the open eye, enabling blood-flow visualization beneath the skin without requiring elaborate immobilization procedures. The openly available recordings support further investigation of blood-flow dynamics beneath the skin and their potential value for vascular monitoring.

\par\medskip
\noindent\textbf{Data and code.}
Raw holograms and media visualizations are available on Zenodo at \href{https://doi.org/10.5281/zenodo.23990028}{doi:10.5281/zenodo.23990028}~\cite{SkinRawDataset2026}. Holovibes v14.14.1 is available from its versioned GitHub repository~\cite{Holovibes14141}.

\bibliographystyle{apsrev4-2}
\bibliography{skin}
\end{document}